\documentclass[twocolumn,showpacs,preprintnumbers,amsmath,amssymb]{revtex4}

\usepackage{graphicx}
\usepackage{dcolumn}
\usepackage{bm}

\begin{document}
\title{Giant Nernst effect in a Kondo lattice close to a quantum critical point}
\author{R. Bel$^{1}$, K. Behnia$^{1}$, Y. Nakajima$^{2}$, K. Izawa$^{2}$, Y. Matsuda$^{2}$, H. Shishido$^{3}$, R. Settai$^{3}$, Y. Onuki$^{3}$}
 \affiliation{(1)Laboratoire de Physique Quantique(CNRS), ESPCI, 10 Rue de Vauquelin,
75231 Paris, France \\ (2)Institute for Solid State Physics, University of Tokyo, Kashiwanoha, Kashiwa, Chiba 277-8581 Japan \\
(3)Graduate School of Science, Osaka University, Tayonaka, Osaka,
560-0043 Japan}

\date{November 19 2003}

\begin{abstract}
We present a study of Nernst and Seebeck coefficients of the
heavy-fermion superconductor CeCoIn$_{5}$. Below 18 K, concomitant
with a field-dependent Seebeck coefficient, a large sub-linear
Nernst signal emerges with a magnitude drastically exceeding what
is expected for a multi-band Fermi-liquid metal. In the mixed
state, in contrast with all other superconductors studied before,
this signal overwhelms the one associated with the motion of
superconducting vortices. The results point to a hitherto unknown
source of transverse thermoelectricity in strongly interacting
electrons.
\end{abstract}


\maketitle

Since the the discovery of superconductivity in
CeCoIn$_{5}$\cite{petrovic}, this Kondo lattice has attracted much
attention. Unconventional superconductiviy\cite{movshovich,izawa}
in this compound occurs in the vicinity of a nearly-avoided
anti-ferromagnetic order and in a metal displaying a pronounced
non-Fermi liquid
behavior\cite{kim,shishido,sidorov,paglione,bianchi3,nakajima}.
The application of pressure\cite{sidorov} or magnetic
field\cite{paglione,bianchi3} leads to the emergence of a Fermi
liquid. It is now generally believed that proximity to a Quantum
Critical Point (QCP) is the origin of the non-Fermi liquid
behavior in the Heavy Fermion (HF) compounds\cite{stewart} and
that unconventional superconductivity mediated by magnetic
fluctuations may arise in such a context\cite{mathur}.

\begin{figure}
\resizebox{!}{0.7\textwidth}{\includegraphics{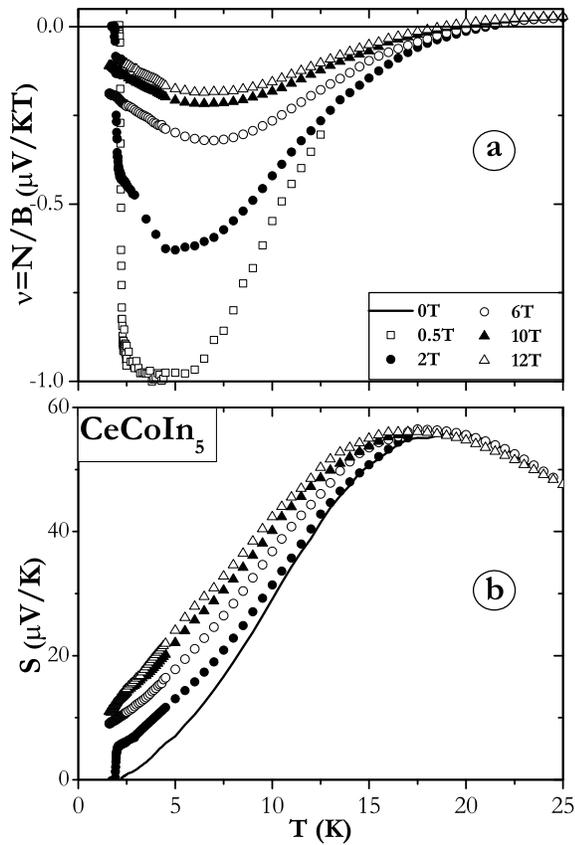}}
\caption{\label{fig1}a) Temperature dependence of the Nernst
coefficient for different magnetic fields. b) Same for the Seebeck
coefficient. Note also the zero-field data for the latter.}
\end{figure}

In this Letter, we present a study of thermoelectric coefficients
in this compound. Unexpectedly, we found a large sub-linear Nernst
signal emerging below 18 K. The magnitude of the Nernst
coefficient in the low-field regime exceeds by far what is
expected in a standard metal.  In this regime, enigmatically, the
ratio of Nernst to Seebeck signals is  diverging with decreasing
temperature and the thermoelectric response of the system tends to
become purely transverse. Moreover, the emergence of this
anomalous Nernst signal is concomitant with a number of non-Fermi
liquid features in various transport properties of the system.
Notably, below 20 K, both resistivity and the inverse of the Hall
coefficient display a linear temperature
dependence\cite{nakajima}. The latter features are archetypes of
non-Fermi liquid transport in cuprates. Our results, by pointing
to quantum criticality as a potential source of enhanced Nernst
effect, may thus offer interesting information for the ongoing
debate on the origin of the anomalous Nernst signal observed in
the normal state of high-T$_{c}$ cuprates\cite{xu,wang1,capan}.

Single crystals of CeCoIn$_{5}$ were grown using a self-flux
method. Thermoelectric coefficients were measured using a
one-heater-two-thermometer set-up. A heat current was injected
into the sample with a small resistive chip. The temperature
gradient created were measured with two Cernox thermometers
attached to local contacts along the sample. Two N11
nanovoltmeters (EM Electronics, UK) were used to measure DC
voltages associated with the longitudinal and transverse electric
fields produced in the sample by the thermal current.

Fig.1 displays the temperature dependence of the Seebeck and
Nernst effects in a single crystal of CeCoIn$_{5}$. Similar
results were obtained in a second sample. As seen in the lower
panel of the figure, the zero-field Seebeck coefficient in this
sample presents a maximum at T$^{*} \sim$ 18 K. Generically, HF
compounds present a maximum in thermopower close to the
temperature which marks the onset of coherent scattering from
Kondo sites\cite{jaccard,gottwick}. Note that here, as in many
other HF systems, the maximum in S(T) occurs at a temperature
which is roughly twice lower than the one associated with maximum
resistivity. As seen in the figure, below T$^{*}$, the zero-field
thermopower decreases rapidly before vanishing in the
superconducting state. The occurrence of superconductivity
apparently impedes a low-temperature sign-reversal observed in
many Ce-based HF systems\cite{link}. The application of a magnetic
field leads to  the enhancement of the Seebeck coefficient below
T$^{*}$. Previous studies of low-temperature thermopower have
detected a strong field dependence in several cases, such as
CeAl$_{3}$\cite{jaccard}, CeRu$_{2}$Si$_{2}$\cite{amato},
UBe$_{13}$\cite{mao} and CeCu$_{6-x}$Au$_{x}$\cite{benz}.
Interestingly, in all these cases, the magnetic field is known to
strongly alter a small characteristic energy of the system. The
strong variation of the Seebeck coefficient with magnetic field in
CeCoIn$_{5}$ can be attributed to a similar effect. At zero-field,
superconductivity occurs before the formation of well-defined
quasi-particles. Thus, $T_{FL}$, the temperature below which the
system displays a Fermi liquid behavior is never attained.  A
field-dependent thermopower may indicate that the magnetic field
significantly increases the energy scale associated with $T_{FL}$.
And indeed, the emergence of Fermi liquid behavior in a moderate
magnetic field\cite{paglione, bianchi3} points to the existence of
such a tuneable energy scale.

As seen in the upper panel of the figure, the transverse
thermoelectricity of the system displays a more striking behavior.
The Nernst signal which is small, positive and field-linear above
18 K, becomes negative, large and strongly non-linear below this
temperature. By plotting the Nernst coefficient $\nu$, defined as
the ratio of transverse electric field to longitudinal thermal
gradient divided by magnetic field ($\nu = N/B = E_{y}/(\nabla_{x}
T B)$), as a function of temperature, this spectacular change of
regime is easily appreciated. The anomalous Nernst signal presents
a broad maximum around a field-dependent temperature which is
about 4 K at 0.5 T and increases to 7 K at 12 T. Note also the
presence of the superconducting transition which leads to a
vanishing Nernst signal in the 0.5 T and 2 T data.

The maximal Nernst coefficient for B=0.5 T is remarkably
large($\sim 1 \mu V / K T$). Its magnitude is comparable with the
maximum attained in the superconducting state of  high-$T_{c}$
superconductors at considerably higher temperatures. It is by far
larger than the residual signal observed above T$_{c}$ in the same
compounds and attributed to vortex-like
excitations\cite{xu,wang1,capan}. No other metal is known to host
a Nernst coefficient of this size with the notable exception of
the giant signal recently discovered in a Bechgaard salt for
fields oriented along the so-called magic angles\cite{wu}.

\begin{figure}
\resizebox{!}{0.7\textwidth}{\includegraphics{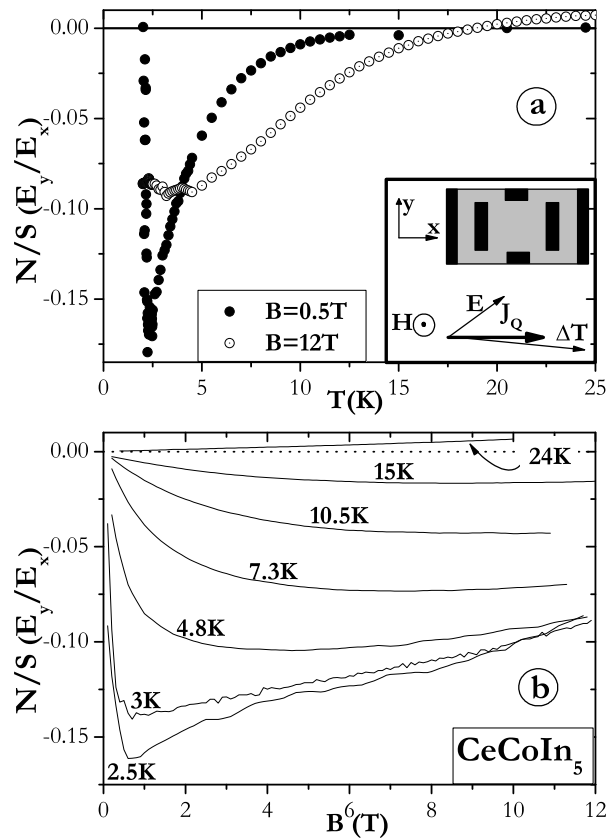}}
\caption{\label{fig2}The ratio of transverse to longitudinal
electric fields produced by the thermal current as a function of
temperature (a) and magnetic field (b). This ratio represents the
tangent of the angle between the longitudinal thermal current and
the induced electric field,$\tan \theta_{TE}$. The inset in (a)
represents the sample geometry and miscellaneous vectors.}
\end{figure}

The very unusual thermoelectricity of CeCoIn$_{5}$ is emphasized
in Fig.2 which presents the ratio of Nernst to Seebeck signals as
a function of temperature and magnetic field. This ratio
represents the tangent of the angle between the thermal current
and the electric field, $N/S = \tan \theta_{TE}$. As seen in the
lower panel of the figure, above T$^{*}$, it is a simple linear
function of magnetic field. A remarkable non-monotonous behavior
emerges when the sample is cooled down below this temperature. In
particular, at lower temperatures, it rapidly becomes a decreasing
function of magnetic field. In other words, \emph{the misalignment
of the electric field decreases with increasing magnetic field.}
As seen in the upper panel of the figure, in the low-field limit
(B=0.5T), $\tan \theta_{TE}$ is a strongly diverging function of
temperature and its enhancement is only interrupted by the
superconducting transition. This divergence is wiped out by the
application of a moderate magnetic field and one observes a
saturation of the ratio at low temperatures. Amazingly, in the
zero-field limit, as the system is cooled down towards T=0,
\emph{the electric field produced by a longitudinal heat current
tends to become purely transverse.}

Now, we turn our attention to the superconducting state. Fig. 3
displays the temperature dependence of the thermoelectric
coefficients in the vicinity of the superconducting transition for
various magnetic fields. As seen in the figure, within the entry
of the system in to the mixed state, both thermoelectric
coefficients of the system decrease monotonously towards zero.
Since below a temperature corresponding to the solidification of
the vortex lattice, one would not expect any finite electric field
in a superconductor, the vanishing of both Nernst and Seebeck
effect is naturally expected. What is striking, on the other hand,
is the absence of any easily detectable Nernst signal associated
with vortex movement under the influence of the thermal gradient.
In the whole range of study, the Nernst signal remains negative
and displays a magnitude smaller than the Seebeck coefficient.
This is in sharp contrast with any superconductor studied until
now. In the mixed state, the thermoelectric response of the system
has been always found to be dominated by the vortex movement along
the thermal gradient due to the entropy carried by the vortex
core. In CeCoIn$_{5}$, this signal is buried under the weight of
the anomalously large signal of the normal state. This can be
shown by comparing the temperature-dependence of Nernst and
Seebeck coefficients, we found that for each field, the
superconducting transition leads to a faster collapse of the
transverse signal. Subtracting a constant fraction of the Seebeck
signal from the Nernst signal, one obtains the vortex-induced
change in the thermoelectric response. As seen in the lower panel
of the figure, the signal thus extracted is finite and positive in
a very narrow temperature window. Comparison  with the resistive
transition in the same magnetic field confirms that the the
extracted signal represents the vortex contribution to the overall
Nernst signal. The magnitude of this contribution is comparable to
what was observed in NbSe$_{2}$\cite{bel} a conventional
superconductor with T$_c$=7.2 K .

As stated above, in CeCoIn$_{5}$, superconductivity occurs before
the formation of well-defined quasi-particles of a Fermi liquid.
It has been recently reported that above a threshold field ($\sim$
5 T) and below a field-dependent characteristic temperature,
T$_{FL}$(H), two signatures of a Fermi liquid are recovered:
resistivity presents a quadratic temperature
dependence\cite{paglione} and the linear term of specific heat,
instead of displaying a logarithmic divergence, tends to a
constant value\cite{bianchi3}. While our study of
thermoelectricity is confined to an area of (H,T) plane which has
a very small overlap with the region of Fermi liquid recovery, it
presents features which support the picture of a field-induced
Fermi liquid.  Above a temperature-dependent threshold field,
close to the T$_{FL}$(H) line, the Seebeck coefficient becomes
almost field-independent and temperature-linear. Moreover, the
magnitude of $S/T$  yields an electronic entropy in very good
agreement with the magnitude of the high-field linear term in
specific heat. Indeed, multiplying $S/T$, by the Avogadro number
and the charge of electron, one obtains $N_{av}eS/T =0.65 J/mol
K^2$  at 12 T, surprisingly close to $C/T= 0.6 J/mol K^2$ reported
at 9 T\cite{bianchi3}. Note, however, that in this Fermi liquid
regime, the Nernst signal, while ceasing to increase as a function
of magnetic field, is still sizeable. It is also important to
notice that the anomalous thermoelectricity reported here is a
transport phenomenon. Indeed, contrary to what has been
theoretically predicted\cite{paul} the zero-field Seebeck
coefficient does not simply follow the temperature-dependence of
the specific heat $C$. While $C/T$ presents a logarithmic
divergence, $S/T$ decreases rapidly with decreasing temperature.
Moreover, for temperatures above $T_{FL}(H)$, the application of
the magnetic field leaves the specific heat virtually
unchanged\cite{bianchi3} but drastically enhances the thermopower.
Finally, we notice that the anomalous transverse thermoelectricity
is sharpest in the zero-field limit and not at B= 5 T which is
supposed to host a QCP.

Explaining the large magnitude of the transverse thermoelectric
response in CeCoIn$_{5}$ is a challenge to the theory. In a
single-band metal, the Nernst coefficient is expected to be
negligible\cite{wang1}. The presence of carriers of different
signs leads to an enhanced signal, known as the ambipolar Nernst
effect\cite{delves}, as recently illustrated in the case of
NbSe$_{2}$\cite{bel}. In the latter system, at T$\sim$ 20 K the
contributions of holes and electrons to the Hall coefficient
cancel out. Such an accidentally compensated metal is expected to
yield a significant ambipolar Nernst effect and indeed the maximum
Nernst signal (0.15 $\mu V / K T$) occurs at this temperature
corresponding to a zero Hall coefficient. In CeCoIn$_{5}$, the
finite field-linear Nernst signal above 18 K is probably linked to
the presence of several bands and carriers of both signs in the
system\cite{shishido}. On the other hand, the large and strongly
non-linear Nernst signal which emerges below this temperature
remains enigmatic. The Hall coefficient in this compound, while
displaying a non-trivial temperature and field-dependence, never
becomes zero\cite{nakajima}. More importantly, the electric Hall
angle does not display any trace of the very peculiar dependence
of $\tan \theta_{TE}$ on magnetic field and temperature.

\begin{figure}
\resizebox{!}{0.7\textwidth}{\includegraphics{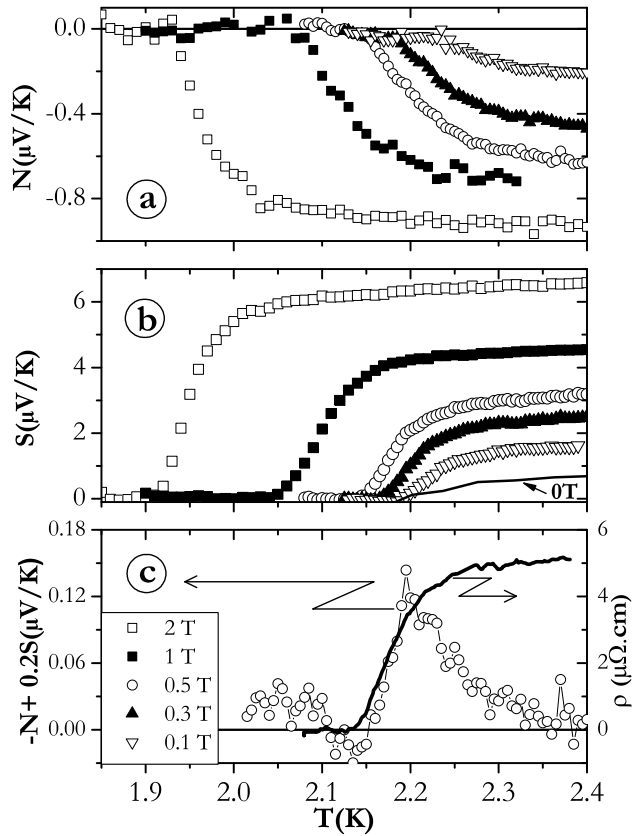}}
\caption{\label{fig3} Temperature dependence of the Nernst
signal(a) and the Seebeck coefficient(b) for different magnetic
fields in the vicinity of the superconducting transition. In panel
(c) the difference between the Nernst signal and a fraction of the
Seebeck signal at 0.5 T is plotted vs. temperature and compared
with the temperature dependence of the resistivity (solid line)
measured in the same conditions.}
\end{figure}

In absence of systematic studies of the Nernst coefficient in
other HF systems, one may speculate in several directions. We
begin by noting that beyond the Relaxation Time Approximation,
anti-ferromagnetic fluctuations may lead to  an enhanced Nernst
signal as well as other non-Fermi liquid transport
properties\cite{kontani}. The possible relevance of vertex
corrections to the anomalous thermoelectricity observed here
remains an interesting direction of investigation.

Another possibility is that the anomalous Nernst signal is a
consequence of the proximity of a QCP. At this stage, no direct
connection between quantum criticality and Nernst effect has been
suggested. However, the behavior of the Hall coefficient in the
vicinity of a QCP has been a subject of
debate\cite{coleman,norman}. A  sudden change in the volume of the
Fermi surface and consequently a sharp anomaly in the magnitude of
the Hall coefficient may occur at the QCP\cite{coleman}. Now, the
Nernst coefficient is intimately related to the off-diagonal
element of the Peltier conductivity tensor $\alpha_{xy}$. The
latter is the energy-derivative of the off-diagonal conductivity
tensor, $\sigma_{xy}$, at the Fermi energy\cite{wang1}:

\begin{equation}\label{1}
\alpha_{xy} =\frac{(\pi k_{B})^{2}T}{3}\frac{\partial \sigma_{xy}
}{\partial \epsilon }\mid _{\varepsilon =\varepsilon _{F}}
\end{equation}
Thus, if, in the vicinity of a QCP, $\sigma_{xy}$ becomes
particularly sensitive  to any slight modification of the volume
of the Fermi surface, this will lead to an enhanced $\alpha_{xy}$
and a sizeable Nernst signal.

An even more exotic scenario is to invoke the presence of
collective modes in a Kondo lattice\cite{flouquet}.
Superconducting vortices are the only mesoscopic objects known to
produce a purely transverse electric field in presence of a
longitudinal thermal gradient. This is because they are entropy
reservoirs associated to a topological defect in a phase-coherent
environment. Very recently, the Kondo lattice problem has been
reconsidered in a two fluid picture with a fluid of heavy
electrons (the condensate) coexisting with a normal fluid of Kondo
impurities\cite{nakatsuji}. The possible existence of entropy
reservoirs associated with orbital moments in such a picture may
constitute a new source of Nernst signal in a manner analogous to
the thermally-induced motion of superconducting vortices.

Future experiments on other HF compounds should tell if an
enhanced Nernst effect is a generic feature of a Kondo lattice.
Clearly, time has come for a serious theoretical treatment of the
problem of the thermoelectric response in a Kondo Lattice close to
a QCP.

In summary, we measured the thermoelectric coefficients of
CeCoIn$_{5}$ and found an anomalous Nernst signal pointing to an
enigmatic source of thermoelectricity in exotic metals.

We are grateful to J. Flouquet, D. Jaccard, H. Kontani, M. Norman
and C. Pepin for their valuable comments.

\end{document}